\documentclass[%
 aip,
 sd,%
 amsmath,amssymb,
 reprint,%
]{revtex4-1}

\usepackage{graphicx}
\usepackage{dcolumn}
\usepackage{bm}

\begin{document}

\preprint{AIP}

\title{ Oblique instability and electron acceleration in relativistic unmagnetized cloud-plasma interaction}

\author{Kazem Ardaneh}
\affiliation{FEMTO-ST Institute, University of Franche-Comte, 25030 Besancon, France}
\email{kazem.arrdaneh@gmail.com}

\date{\today}

\begin{abstract}
Relativistic unmagnetized cloud-plasma interaction is analyzed by performing linear analysis and particle-in-cell simulation. This course consists of an electron-ion cloud injected into a stationary ambient plasma and has long been a favorite topic in laboratory and space plasmas. An oblique electromagnetic instability dominates the unstable spectrum. In the interaction with the generated electromagnetic fields, the cloud electrons are entirely mixed with the ambient ones and form a hot electron population. The velocity of the cloud ions, however, has not changed significantly from the initial bulk velocity. As this ion cloud propagates into the plasma, it derives an electrostatic field which can accelerate the electrons up to energy equipartition between electrons and ions. The electrostatic field is amplified at the expense of the kinetic energy of ions, and its spatial scale is in order of the electron skin depth. The electron acceleration in such an electrostatic field is, therefore, a likely process for pre-acceleration of electrons in unmagnetized plasmas.

\end{abstract}

\maketitle

\section{\label{Introduction}Introduction}
Propagation of a relativistic stream of plasma into an ambient plasma exists in a variety of astrophysical and engineering systems, e.g., gamma-ray bursts, and active galactic nuclei jets in astrophysics and fast ignition scenario to compress and heat the fuel in the inertial confinement fusion. Several plasma instabilities are involved in the relativistic unmagnetized cloud-ambient interaction, e.g., electrostatic (ES) two-stream or Buneman \citep{bun58} instability and electromagnetic (EM) filamentation  \citep{fri59} or Weibel instability \citep{wei59}. The nature of instability is generally an oblique mode \citep{brd10,ard15,ard16}. Depend upon the system parameters, one of the modes might dominate the whole unstable spectrum.

The formation of collisionless shocks and acceleration of particles are ubiquitous in the relativistic cloud-ambient interaction \citep{ken09,ard15,ard16,cho14,yao18}. The shock accelerated electrons are believed to be the origin of the prompt and afterglow missions of gamma-ray bursts \citep{don17,men18}. The diffusive shock acceleration (DSA) is the widely accepted scenario for acceleration of particles in collisionless shocks \citep{bel78,bel13,blan78,blan87,spi08,mar09,sir11,sir13} in which particles attain energy while they oscillate around the shock front as the result of scattering back and forth by MHD waves. 

Only those particles with energies well beyond their thermal ones are qualified for acceleration via DSA. In the absence of any pre-acceleration, thermal electrons which are closely tied to magnetic field lines convect downstream without undergoing remarkable acceleration. It is not well understood how electrons might reach the threshold energy of DSA. The lack of a fully self-consistent theory of electron pre-acceleration is referred to as electron injection problem.

The motional electric field of the magnetized clouds, {\boldmath${E_{0}}=-{\beta_{0}}\times{B_{0}}$}, might straightly inject electrons into the DSA. The gradient of the magnetic field in the transition region of the shock can drift electrons perpendicular to the shock propagation direction and capable them for acceleration by the motional electric field. This process is referred to as shock drift acceleration (SDA)  \citep{web83,beg90,par12,par13,guo14,li17}. On another hand, in the magnetized electron-ion clouds, the ES waves in the transition region of the shock generated by an upper-hybrid wave instability might trap the electrons. They can then be accelerated by the motional electric fields while surfing in the perpendicular direction. This process is known as electron surfing acceleration (ESA)  \citep{hos01,hos02,li18}.

In relativistic unmagnetized cloud-plasma interaction, we expect an alternative process because {\boldmath${E_{0}}=0$}. The plasma instabilities in the relativistic regime are generally EM and therefore strong low-frequency EM waves are emitted. The electrons can be effectively heated in interaction with these waves. There is another important component to be addressed. Because electrons are easily affected by the magnetic field lines while ions are not, deceleration of the cloud ions takes much longer time. As the cloud ions propagate into the ambient plasma, an ES field arises which might accelerate the electrons up to energy equipartition between electrons and ions.

The main objective of the current work is electron pre-acceleration in relativistic unmagnetized cloud-plasma interaction where collisionless shocks are formed as the relativistic cloud propagates into the ambient plasma.  In this paper, we have analyzed this course using linear analysis and Particle-In-Cell (PIC) simulation. Based on the linear analysis, propagation of a cold cloud into a cold ambient plasma excites oblique modes. Furthermore, the cloud ions derive another oblique mode where its ES component is more pronounced. The results of the PIC simulation show that at early times, strong transverse EM fields generated by an oblique instability heat the electrons in transverse directions which results in symmetric phase-space distribution.  At later times, the ES field accelerates the electrons at the expense of the kinetic energy of the ions.

This paper is structured as follows. The linear analysis is presented in Section \ref{Linear analysis}. The PIC simulation setup is presented in Section \ref{PIC simulation}. The results of the simulation are disscused in Section \ref{Simulation results}. We conclude with a summary in Section \ref{Summary and conclusions}. Dimensionless groups are used throughout this paper where space is normalized to $c/\omega_{\rm pe}$, time to $\omega_{\rm pe}$, mass to $m_{\rm e}$, charge to $e$, velocity to $c$, momentum to $m_{\rm e}c$, electric and magnetic fields, in the CGS system of units, to $m_{\rm e}\,c\,\omega_{\rm pe}/e$, vector potential to $m_{\rm e}\,c^2/e$, energy density of electric and magnetic fields to $\sum_{\rm s}(\Gamma_{\rm s}-1)n_{\rm s}m_{\rm s}c^2$, and density to the ambient density $n_{\rm a}$. Here $s$ denotes the $s$th specie (electron or ion). The dimensionless form of a quantity, e.g., $x$, is shown as $x^{*}$.

\section{Linear analysis}\label{Linear analysis}

For a collisionless plasma, the distribution function $f_{\rm s}$ of each species evolves according to the relativistic Vlasov equation. Separating the distribution function into an unperturbed part and an infinitesimal perturbation, $f_{\rm{s}}= f_{s0} +\delta f_{\rm{s}}$,  and considering the unperturbed part spatially uniform,  $f_{\rm{s0}}({\bf r^{*}},{\bf p^{*}})=n^{*}_{\rm{s}}\varphi_{\rm{s}}({\bf p^{*}})$, one can obtain the following dispersion relation \citep{brd10}:

\begin{equation} \label {dispersion}
( \omega^{*2}D_{\rm xx} - k^{*2}_{\rm y} )(\omega^{*2}D_{\rm yy}-k^{*2}_{\rm x})-(\omega^{*2}D_{\rm xy}+k^{*}_{\rm x}k^{*}_{\rm y})^2=0
\end{equation}

where 

\begin{multline} \label {dis}
D_{\rm mn}=\delta_{\rm mn}- i{\sum_{\rm{s}}}\frac{1}{m^{*}_{\rm s}\omega^{*2}}\int_{0}^{\infty}d\tau^{*}\int d^3{\bf p^{*}}\\
\times\exp[i(\gamma\omega^{*}-{\bf k^{*}}.{\bf p^{*}})\tau^{*}]\\
\times p^{*}_{\rm m}[({\omega^{*}-\frac{{\bf k^{*}}.{\bf p^{*}}}{\gamma}})\frac{\partial}{\partial p^{*}_{\rm{n}}}
+\frac{p^{*}_{\rm{n}}}{\gamma}{\bf k^{*}}.{\nabla_{\rm p^{*}}}]\varphi_{\rm{s}}({\bf p^{*}})
\end{multline}

The wave vector is assumed in $xy$-plane, ${\bf k^{*}}=(k^{*}_{\rm x}, k^{*}_{\rm y})$. Let us assume that the cloud initially move along the $x$-direction with momentum $p^{*}_0$ and thermal spread $p^{*}_{\parallel}$, and $p^{*}_{\perp}$ along the $x$- and $y$-direction. It is considered cold along the $z$-direction. The ambient plasma is assumed cold. We use the so-called water-bag distribution function given by 

\begin{multline} \label {PDF}
\varphi({\bf p^{*}})=\frac{1}{4p^{*}_{\parallel}p^{*}_{\perp}}\delta(p^{*}_{\rm{z}})[H(p^{*}_{\rm{y}}+p^{*}_{\perp})-H(p^{*}_{\rm{y}}-p^{*}_{\perp})]\\
\times[H(p^{*}_{\rm{x}}+p^{*}_{\parallel}-p^{*}_0)-H(p^{*}_{\rm{x}}-p^{*}_{\parallel}-p^{*}_0)]
\end{multline}

where $H(x-a)$ is the Heaviside step function with $H(x < a)=0$ and $H(x\geq a)=1$, and $\delta(x)$ is the delta function. The expression for each $D_{\rm mn}$ in Eq. \ref {dispersion} and dispersion relation can be obtained analytically after some straightforward but lengthy algebra. The $D_{\rm mn}$ elements are given in the Appendix \ref{Tensor elements for water-bag distribution functions}. Setting $p^{*}_{\parallel}=p^{*}_{\perp}=0$, one can obtain the  $D_{\rm mn}$ elements for a cold electron-ion cloud propagating into a cold ambient plasma.

\begin{subequations} \label {cold}
\begin{align}
D_{\rm xx}=1-\frac{1}{\mu\omega^{*2}}-\frac{n^{*}_{\rm c0}}{\mu\omega^{*2}\gamma^{3}_{0}}\frac{\omega^{*2}+k^{*2}_{\rm y}p^{*2}_{0}}{(\omega^{*}-k^{*}_{\rm x}\beta_{0})^2} \\
D_{\rm yy}=1-\frac{1}{\mu\omega^{*2}}-\frac{ n^{*}_{\rm c0}}{\mu\omega^{*2}\gamma_{0}}\\
D_{\rm xy}=-\frac{n^{*}_{\rm c0} k^{*}_{\rm y}\beta_{0}}{\gamma_{0}\mu\omega^{*2}(\omega^{*}-k^{*}_{\rm x}\beta_{0})} 
\end{align}
\end{subequations}
 
where $\mu=R/(R+1)$, and $R=m_{\rm i}/m_{\rm e}$. Several plasma instabilities might be excited in the relativistic cloud-plasma interactions: (1) purely transverse EM mode with a wave vector perpendicular to the cloud propagation direction (filamentation mode),  (2) EM instability driven by thermal anisotropy with a wave vector along the lower temperature axis (Weibel mode), and (3) electrostatic modes such as two-stream and Buneman  instabilities with a wave vector parallel to cloud propagation direction. 

\begin{figure}
\includegraphics[scale=0.4]{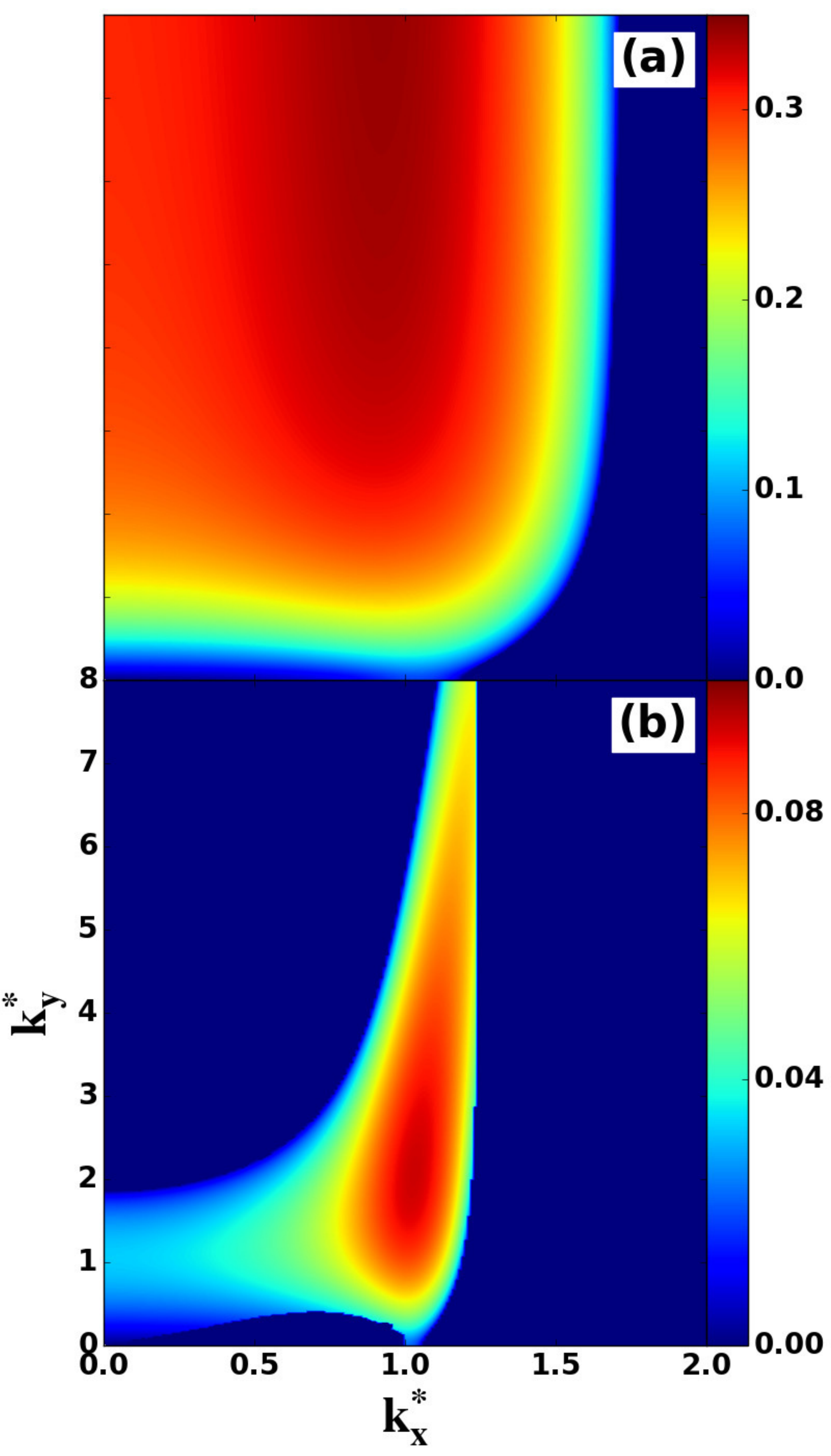}
\caption{ Growth rate map for (a) a cold cloud propagating into a cold ambient plasma, (b) a hot ion cloud propagating into a cold ambient electron. For both cases $\gamma_{0}=10$. In panel (b),  $p^{*}_{\parallel}=0.5$, and $p^{*}_{\perp}=0.5$.}
\label{grow_rate_map}
\end{figure}

Considering $k^{*}_{\rm y}=0$ in the Eq. \ref{dispersion}, the instability reduces to the electrostatic beam-plasma instability, with a dominant wavenumber given at $k^{*}_{\max}\simeq \beta_{0}^{-1}$ and a maximum growth rate  $\Im\omega^{*}_{\max}\simeq\frac{1}{2\gamma_0}(\frac{3}{\mu})^{1/2}(\frac{n^{*}_{\rm c0}}{2})^{1/3}$ for the two-stream instability and $\Im\omega^{*}_{\max}\simeq\frac{\sqrt{3}}{2\gamma_0}(\frac{n^{*}_{\rm c0}}{2R})^{1/3}$ for the Buneman instability. The EM beam-plasma instability, on the other hand, can be obtained using $k^{*}_{\rm x}=0$ which results a maximum growth rate  $\Im\omega^{*}_{\max}\simeq\beta_0(\frac{n^{*}_{\rm c0}}{\mu\gamma_0})^{1/2}$ and dominant wavenumber as $k^{*}_{\max}\gg \beta_{0}^{-1}$. The nature of the instability, however, is an oblique instability as both longitudinal and transversal waves components are present at the same time, even in the cold limit.
 
Two stages are considered for the propagation of the relativistic cloud into the ambient plasma. The initial stage is the propagation of a cold electron-ion cloud into an ambient plasma. At a later time, the cloud electrons will be entirely mixed with ambient ones and form a hot electron population which moves with a drift velocity perpendicular to the EM fields. The cloud ions, on the other hand, are slightly heated. Hence, as an intermediate stage,  propagation of a slightly hot ion cloud into an ambient plasma is considered. It is assumed that the cloud particles move with drift velocity $\beta_{\rm c0}=0.995$ (bulk Lorentz factor $\Gamma_{\rm c0}=10$). A thermal spread $(p^{*}_{\parallel}, p^{*}_{\perp})=(0.5, 0.5)$ is considered for the intermediate stage.  

The growth rate maps for the two stages are illustrated in Figure \ref{grow_rate_map}, panel (a) corresponds to the initial stage and panel (b) corresponds to the intermediate stage. As one can see in panel (a), both electrostatic and electromagnetic modes are present for cold cloud propagating into the ambient. Therefore, the oblique modes are dominant for a cold electron-ion cloud propagating into an ambient plasma. This is in contrast to \citet{brd10} where filamentation modes dominate for $n^{*}_{\rm c0}=1$. It is due to the asymmetry between the cloud and ambient in the present study while in \citet{brd10}  the beam and the return current are perfectly symmetric under the assumption of $n^{*}_{\rm c0}=1$ and thus the filamentation modes are dominant. 

The EM fields heat the particles in the transverse direction while electrostatic fields heat the particles in longitudinal direction. As shown in panel (b), the thermal spread reduces the continues unstable modes to a dominant mode at $(k^{*}_{\rm x}, k^{*}_{\rm y})\approx(1, 2 )$, closer to the parallel axis and electrostatic approximation in comparison to the panel (a). It also exhibits a significant suppression of the filamentation instability.  The location of the dominant mode basically depends on the thermal spread and drift velocity of the cloud particles.

\section{PIC simulation}\label{PIC simulation}

\begin{figure*}
\includegraphics[scale=0.5]{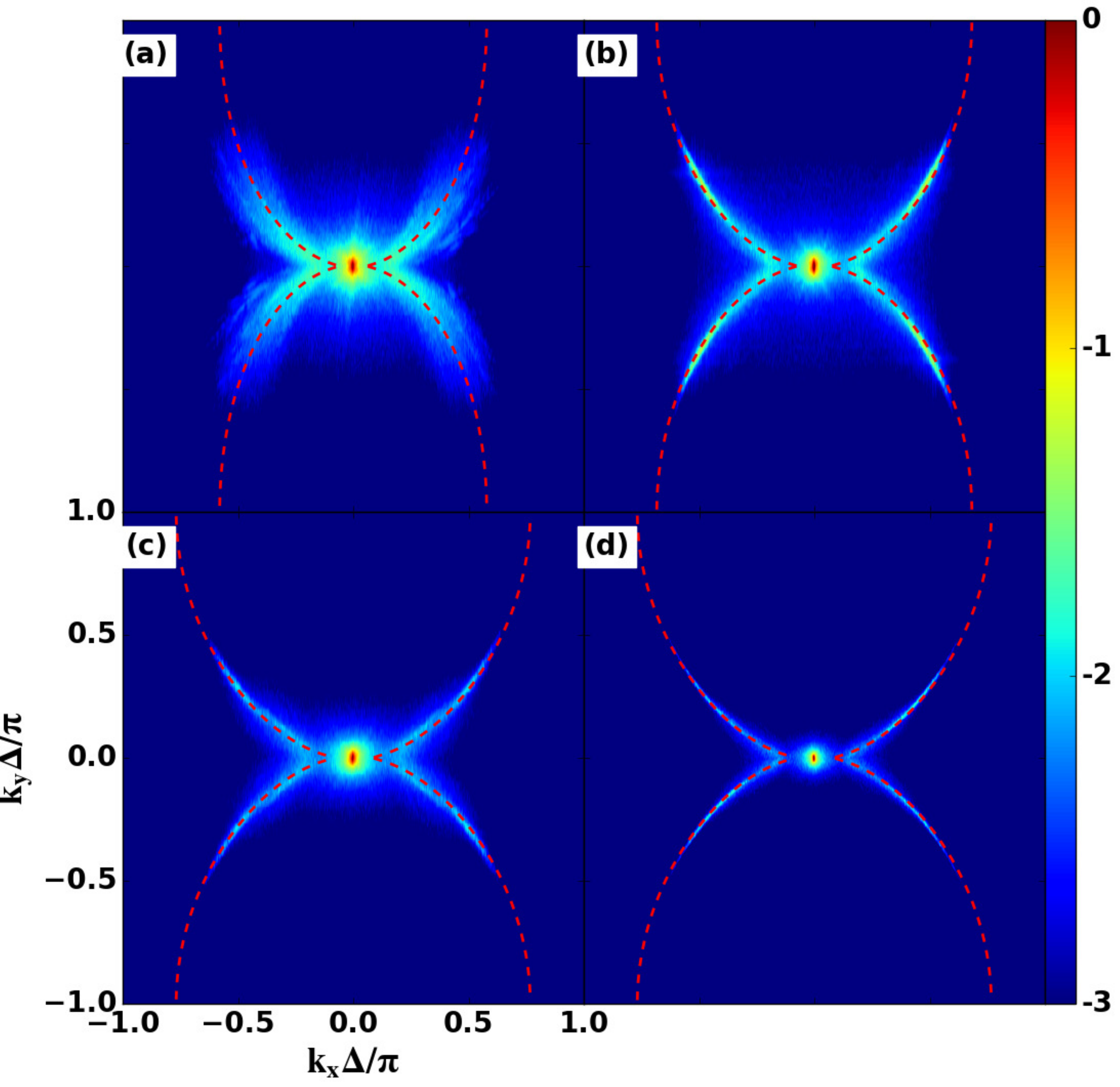}
\caption{ The $z$-component of the magnetic field in Fourier space with a CFL number of (a) 1.0, (b) 0.7, (c) 0.5, and (d) 0.5 at $t^{*} = 600$. The simulation box is identical for panels (a)-(c) with a spatial resolution of $c/\omega_{\rm pe}/\Delta=10$ while for panel (d) $c/\omega_{\rm pe}/\Delta=20$. For all panels, the implicit factor is $\theta=0.501$ and second order shape is used. The color-bar is on a logarithmic scale. In all panels, Eq. \ref{FDTDdispersion2} is plotted as a red dashed line which shows the wavenumber regions of the numerical Cherenkov radiation.}
\label{cherenkov}
\end{figure*}

The code employed in the present work is a modified version of the relativistic EM particle code pCANS developed at Chiba University.  The coupled equations for the particle and fields are solved using an operator split algorithm. The trajectory of each particle is integrated using Buneman-Boris method as discussed in \citet{bir91} while Maxwell's curl equations are solved implicitly using the conjugate gradient method. The code uses CGS system of units. A series of test simulations have already been performed to establish a numerical model which best conserves energy and minimizes numerical self-heating. The simulation is performed on a computational box  $(L^{*}_{\rm x}, L^{*}_{\rm y})=(800.5, 51.2)$. There are 20 particles per cell per species for both cloud and ambient plasmas, $n^{*}_{\rm c0}=1$. The frame of reference is ambient, in which the cloud plasma propagates in positive $x$-direction with bulk Lorentz factor $\Gamma_{\rm c0}=10$. The cloud fills the whole computational domain in the $y$-direction and is injected continuously at $x^{*}_{\rm c0} = 150$. The cloud and ambient plasmas are initially cold and unmagnetized. The ion-to-electron mass ratio is $R=25$. The reflecting boundary is used in the $x$-direction while periodic boundary condition is applied for $y$-direction.  

The numerical Cherenkov radiation is one of the crucial concerns in investigating cold relativistic plasma streams using PIC simulations with the standard Yee finite difference time domain (FDTD) scheme \citep{god74}. In this scheme,  the phase speed of the electromagnetic wave is numerically less than the speed of light in high-wavenumber regions. Several techniques have been proposed for suppressing the numerical Cherenkov radiation, e.g.,  using the higher-order solver for Maxwell's curl equations and applying weak Friedman filter to waves in high-wavenumber regions as presented in  \citet{gre04} and applied in the new versions of TRISTAN PIC code \citep{spi08,sir11,sir13,ard15,ard16}. However, the filters might cause numerical damping of physical waves which is particularly problematic in studying particle accelerations in collisionless shocks. 

For pCANS PIC code, where Maxwell's curl equations are solved implicitly, \citet{ike15} reported that numerical Cherenkov radiation was considerably suppressed with a Courant-Friedrichs-Lewy (${\rm CFL}=c\Delta_{\rm t}/\Delta$) number of 1.0. Additionally, it is found that using second order shape functions and an optimal implicitness factor of $\theta=0.501$ further suppressed long-wavelength modes of the numerical instability. 

Shown in Figure \ref{cherenkov} are four PIC simulations performed using pCANS. The employed CFL numbers are 1.0,  0.7, 0.5, and 0.5 for panels (a)-(d), respectively. The simulation box is identical for panels (a)-(c) with a spatial resolution of $c/\omega_{\rm pe}/\Delta=10$, while for panel (d) $c/\omega_{\rm pe}/\Delta=20$. In all cases, the implicit factor is $\theta=0.501$ and second order shape is used. As one can see, the waves that are excited in high-wavenumber regions are certainly the intersections indicated by Eq. \ref{FDTDdispersion2} (red dashed lines) and typical indications of numerical Cherenkov radiation. Moreover, we notice that numerical Cherenkov radiation gradually suppressed as the CFL reduces from 1.0 to 0.5. It is in contrast with the magical CFL number of 1.0 reported in \citet{ike15}. 
Figure \ref{cherenkov}d shows that the numerical Cherenkov instability is significantly suppressed under ${\rm CFL}=0.5$ and $c/\omega_{\rm pe}/\Delta=20$, which can be simplified as $\omega_{\rm pe}\Delta_{\rm t}=1/40$. In the Section \ref{Simulation results}, the results of this case will be presented.

\section{\label{Simulation results}Simulation results}

\begin{figure}
\includegraphics[scale=0.4]{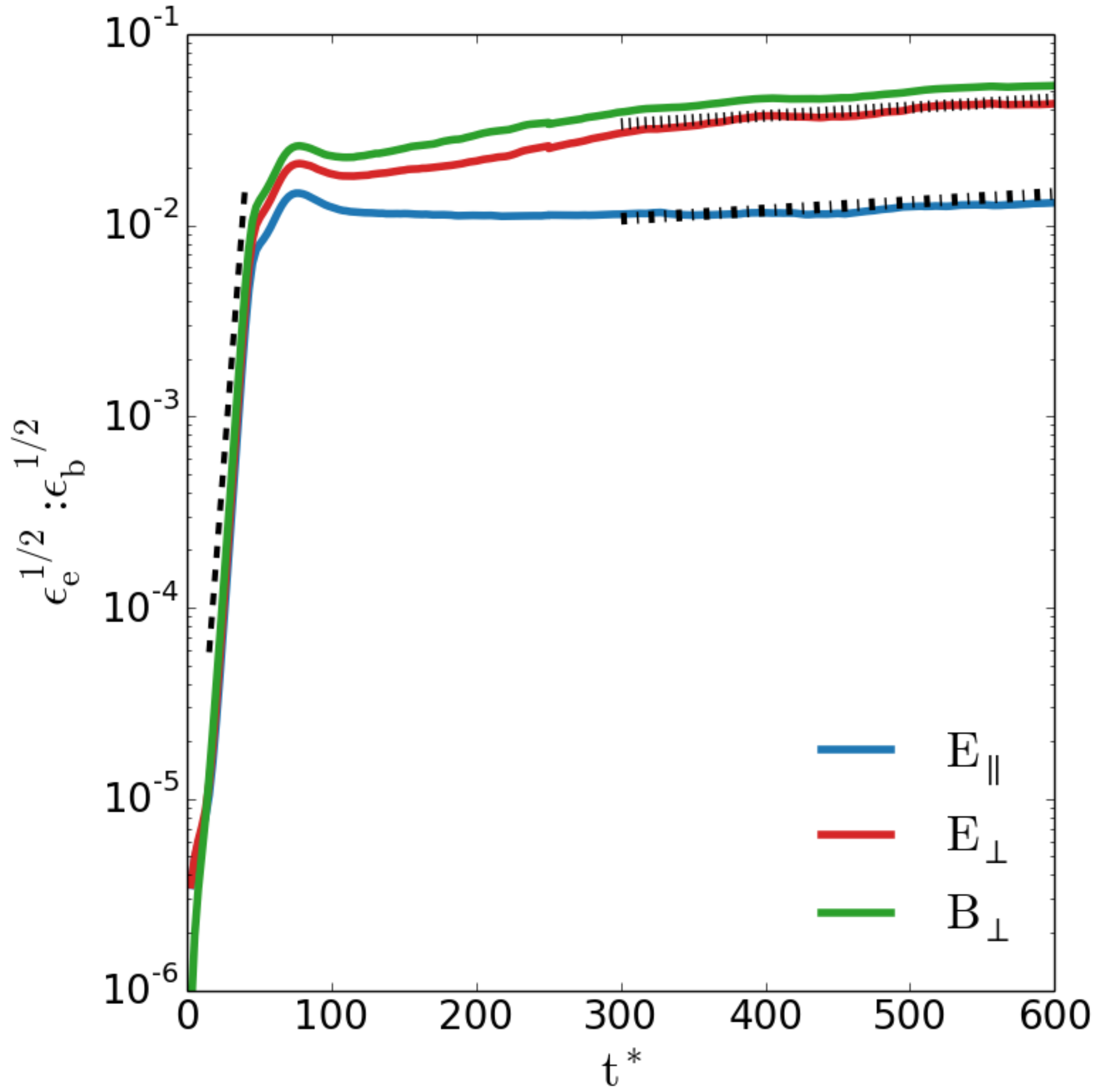}
\caption{Time evolution of energy density for longitudinal electric field (blue line), transverse electric field (red line), and transverse magnetic field (green line). The dashed line shows $\epsilon\propto\exp (0.25\,t^{*})$ while the dot and dashed-dot lines show $\epsilon\propto\exp (0.001\,t^{*})$.}
\label{E_energy}
\end{figure}

The time evolution of the EM-ES fields energy densities shows the involved instabilities in the cloud-ambient interaction. 
The box-averaged energy density of fields is calculated as $e_{\rm F}(t)=\frac{1}{8\pi N_{\rm x}N_{\rm y}} \sum_{\rm {i=1}}^{\rm {N_x}}\sum_{\rm {j=1}}^{\rm {N_y}} F^2(i\Delta,j\Delta,t )$ where $F$ denotes electric or magnetic field. As mention in Section \ref{Introduction}, the energy density of fields is normalized relative to the initial kinetic energy density of cloud which is $\sum_{\rm s=i,e}(\Gamma_{\rm s}-1)n_{\rm s}m_{\rm s}c^2=4680\,m_{\rm e}c^2$ for the current simulation. As shown in Figure \ref{E_energy}, both longitudinal and transversal electric fields grow linearly in the early stage, $0\lesssim t^{*}\lesssim40$ with same growth rate of $\Im\omega^{*}=0.29$. The nature of the instability is, therefore, an oblique instability which has both ES ($E_{\rm x}$) and EM ($E_{\rm y}$) wave components, in agreement with the Figure \ref{grow_rate_map}a in Section \ref{Linear analysis}. This growth is at the expense of the kinetic energy of the incoming cloud electrons. The EM fields then become strong enough to affect the much heavier species of the ambient and the ions start to participate in the instability. Moreover, the particles are heated by the induced fields where the electrons and ions are not thermal equilibrium anymore due to their different inertia. Between $100\lesssim t^{*}\lesssim300$ the growth of longitudinal and transversal electric fields diverges. The electromagnetic component grows faster than the electrostatic one. Therefore, the particles are mainly heated in the transverse direction and the nature of instability is a Weibel-like instability due to the thermal anisotropy in the distribution of the particles \citep{wei59}. The interesting part is the continuous growth of the ES and EM components from $t^{*}\simeq300$ until end of simulation. These fields are associated with an oblique instability as the cloud ions propagate into the ambient plasma (Figure \ref{grow_rate_map}b). The growth of this instability is at the expense of the cloud ions kinetic energy. In the following sections, we will discuss the role of this instability in electron acceleration.

\subsection{Electron acceleration}\label{Electron acceleration}

\begin{figure}
\includegraphics[scale=0.45]{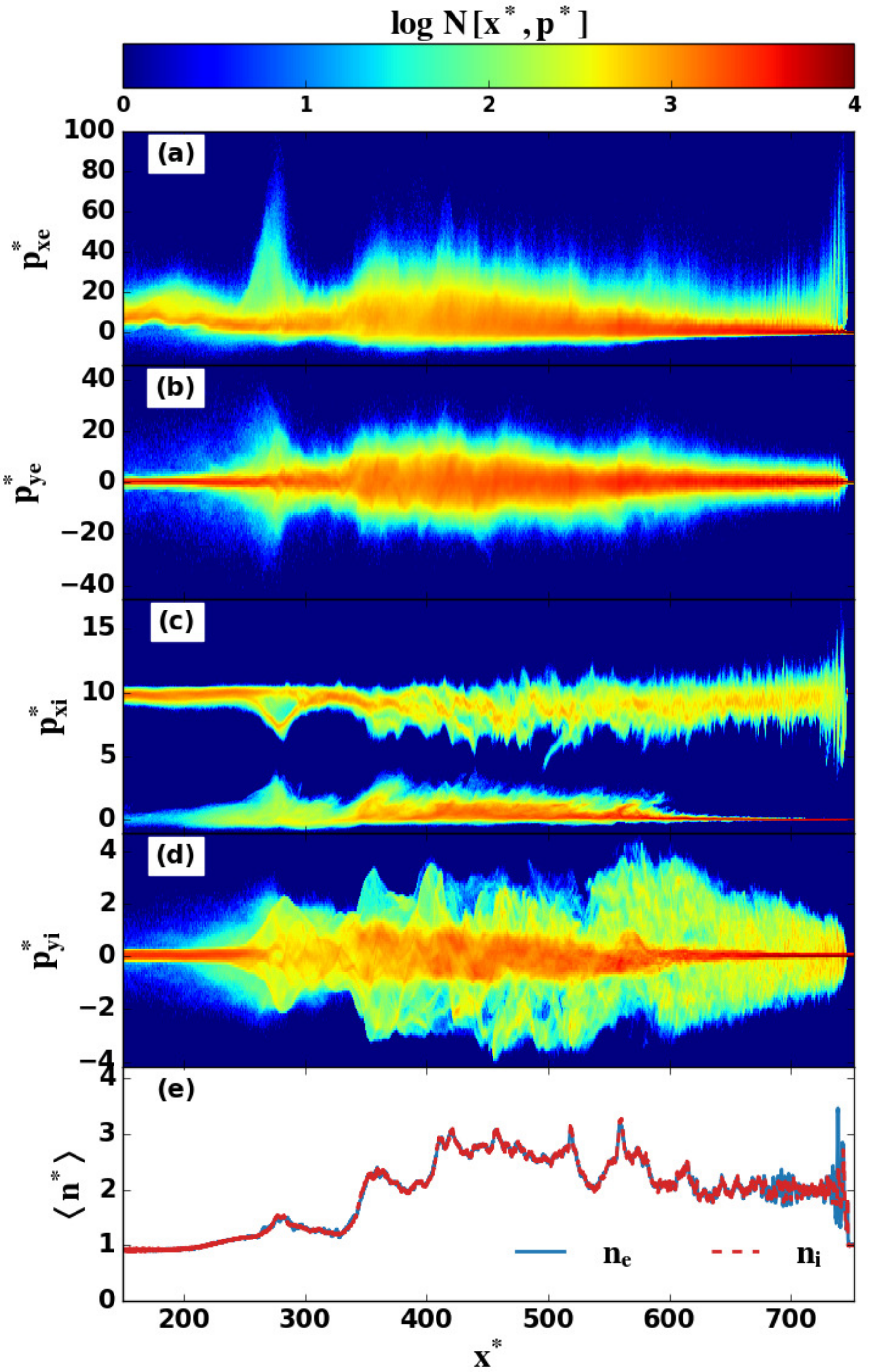}
\caption{ Phase-space distribution of the cloud-ambient interaction at $t^{*}=600$ for: panel (a) longitudinal phase-space distribution of electrons, panel (b) transverse phase-space distribution of electrons, (c) longitudinal phase-space distribution of ions, and panel (d) transverse phase-space distribution of ions. The profiles of the transversely averaged electron, and ion densities are shown in panel (e), blue line for electron and red dashed line for the ion, respectively. Due to the very large number of particles in the simulation, $8\times10^7$ particles are randomly selected in all panels. As particles are selected randomly, the respective distribution function is not affected.}
\label{phase_space}
\end{figure}

Propagation of a relativistic cloud into an ambient medium forms a double shock structure \citep{ken09,ard15,ard16}. Being lighter than ions, at first, the cloud electrons are decelerated while the ambient electrons are swept \citep{ard14,ard16}.  It is shown that for the cloud-to-ambient density ratio greater than one, a double peak will form in the total electron density which is recognized as a double shock system. Contribution of the ions in the shock system is slightly different. When contact discontinuity forms, contact discontinuity is defined as the interface between the trailing shock which forms in the cloud electrons and leading shocks which form in the ambient, a population of the ambient ions is trapped behind the trailing shock which contributes in the trailing shock. Another population of ambient ions located in the right side of the contact discontinuity is swept by the cloud flow and contributes into the leading shock.  The cloud to ambient density ratio is unity in the current study. As a result, we see a compressed region where the shocked to  un-shocked density ratio reaches $\sim3$ (Figure \ref{phase_space}e). Although the could ions are not slowed down and not contributing in the shock system, the ambient ions are shocked and conserve the charge neutrality of the system.  The shocked region is between  $400\lesssim x^{*}\lesssim500$ (Figure \ref{phase_space}e). Both sides of the shock are dominated by the fields generated by oblique instability. There are transverse EM fields as well as a longitudinal ES field (Figure \ref{E_energy}). 

The electrons might attain energy in interaction with transverse EM fields. In this interaction, the phase-space of the particle is symmetric (Figure \ref{phase_space}b and \ref{phase_space}d) because the transverse components of generalized momentum are conserved $p^{*}_{\rm y}+q^{*}/m^{*}A^{*}={\rm const}$. The phase-space of the electron and ion are correlated to each other as $p^{*}_{\rm yi} \simeq p^{*}_{\rm ye}/R$ (Figure \ref{phase_space}b and \ref{phase_space}d). As one can see, the cloud and ambient electrons are entirely mixed and indistinguishable (Figure \ref{phase_space}a). However, the cloud ions move with their initial bulk velocity and are distinguishable from the ambient ones. Propagation of this ion cloud into the ambient plasma results in an oblique instability as discussed in Section \ref{Linear analysis}. 

\begin{figure}
\includegraphics[scale=0.4]{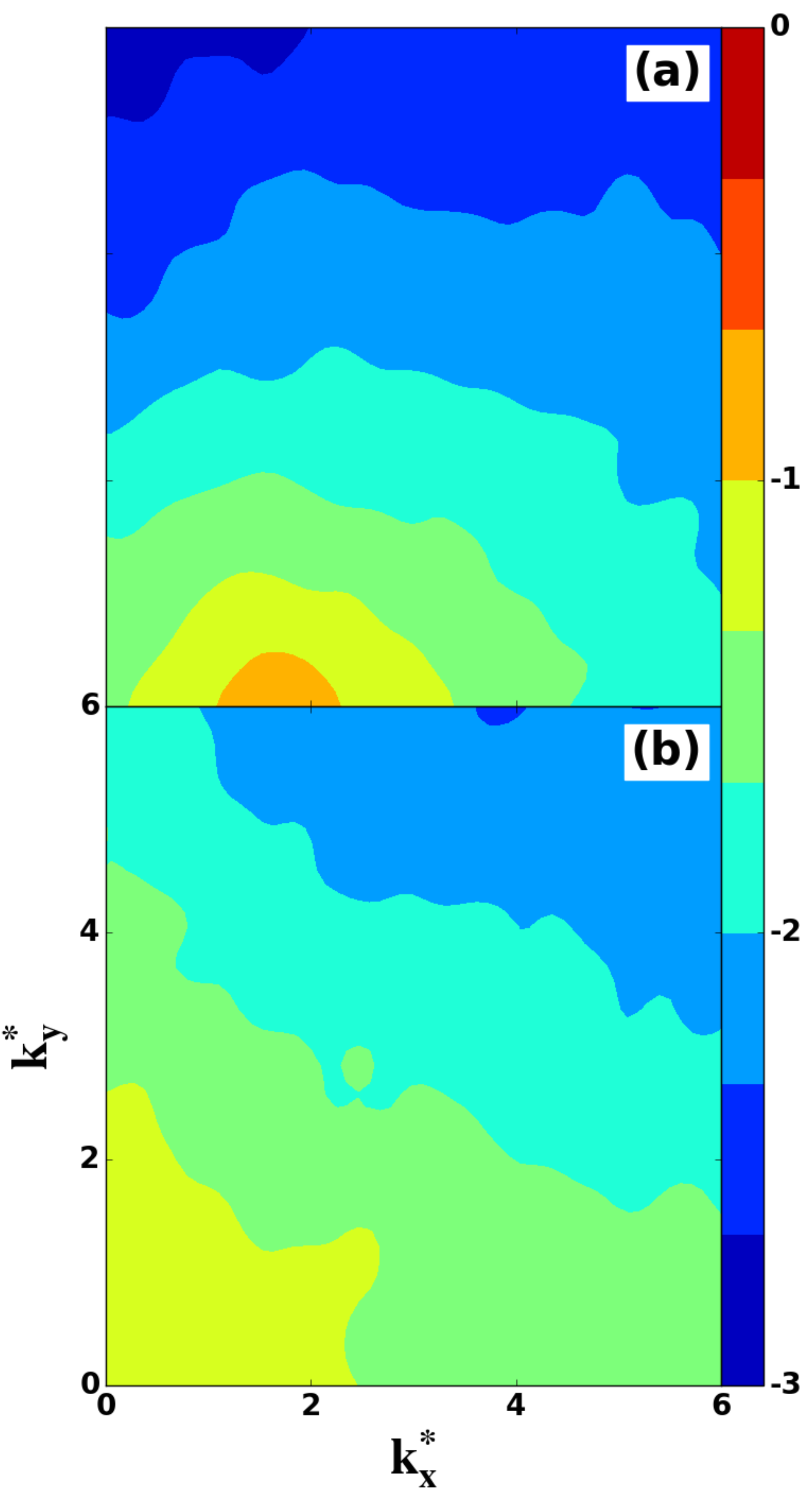}
\caption{The $x$-component, panel (a), and $y$-component, panel (b), of the electric field in Fourier space. The Fourier transfer is performed for the interested region $-100\leq x^{*}-\beta_{\rm c0}t^{*}\leq0$. The color-bar is on a logarithmic scale.}
\label{FFTExEy}
\end{figure}

Let us look over the EM-ES fields excited due to the propagation of the ion cloud into the ambient plasma. We zoom in the region $-100\leq x^{*}-\beta_{\rm c0}t^{*}\leq0$ at $t^{*}=600$ and calculate the Fourier transfer of the $x$-component, and $y$-component of the electric field as shown in Figure \ref{FFTExEy}.
We have shown in Figure \ref{grow_rate_map}b that propagation of a slightly hot beam ion with thermal spread of $(p^{*}_{\parallel}, p^{*}_{\perp})=(0.5, 0.5)$, and drift moment of $p_{\rm c0}=10$ into a plasma reduces the continues unstable modes to a dominant mode closer to the parallel axis and electrostatic approximation. Figure \ref{FFTExEy} demonstrates a similar behavior for the EM-ES fields within the region $-100\leq x^{*}-\beta_{\rm c0}t^{*}\leq0$ where the ES component of the instability is more pronounced compared to the EM one. Hence, the electron heating in the transverse direction ($\vert p^{*}_{\rm ye}\vert \sim 5$) is much smaller than the acceleration of the electrons in the $x$-direction ($p^{*}_{\rm xe} \sim 20$), see region $x^{*}\gtrsim 650$ in Figure \ref{phase_space}a and \ref{phase_space}b.

\begin{figure*}
\includegraphics[scale=0.4]{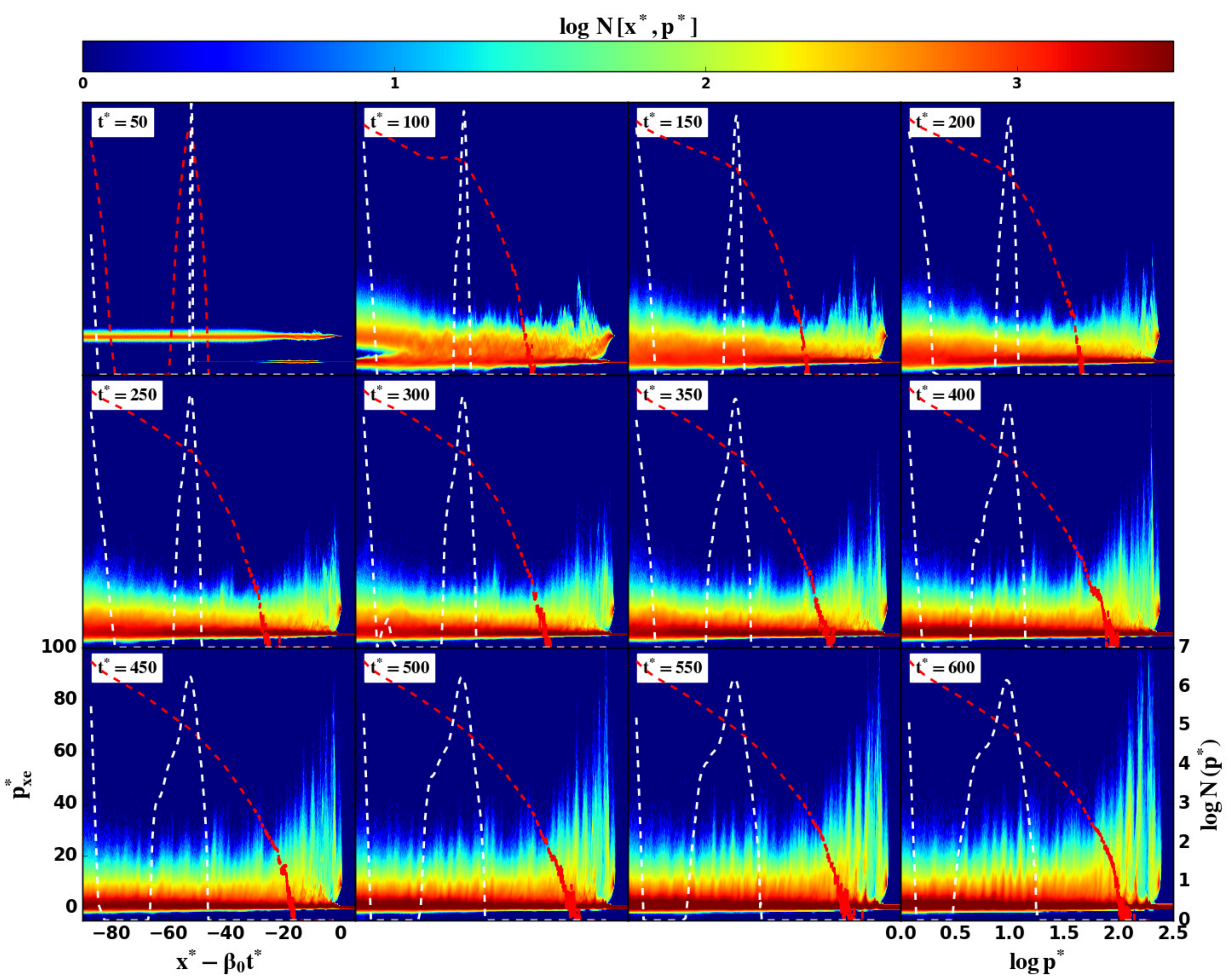}
\caption{Evolution of the phase-space distribution of electrons between $-100\leq x^{*}-\beta_{\rm c0}t^{*}\leq0$. The position $x^{*}$ is measured from the cloud front. The over-plotted lines in each panel show electron spectrum (red dashed line), and ion spectrum (white dashed line). The axes of the phase-space plots are given in bottom left, while the axes of the particle spectrum plots are given in the bottom right.}
\label{phase_space2}
\end{figure*}

The amplification of the ES field continues until the energy equipartition between electrons and ions. To demonstrate the acceleration more clear, let us focus on the time evolution of electron phase-space within the $-100\leq x^{*}-\beta_{\rm c0}t^{*}\leq0$ (Figure \ref{phase_space2}). As one can see, the energy of electrons increases as the amplitude of the ES field grows. At the same time the cloud ions are decelerated and their kinetic energy is consumed for acceleration of the electrons. The efficiency of energy exchange between the cloud ions and electrons is twice the growth rate of the  Buneman instability, $\simeq\frac{\sqrt{3}}{\gamma_0}(\frac{n^{*}_{\rm c0}}{2R})^{1/3}$. It indicates that the energy exchange becomes inefficient as the ion-to-electron mass ratio approaches the realistic value. The wavelength of the electron oscillation (inset in Figure \ref{phase_space2}) is approximately wavelength of the ES field, $\lambda^{*}\sim 1$ (Figure \ref{grow_rate_map}b). As the result, there is resonant electron acceleration. 

The over-plotted lines in Figure \ref{phase_space2} are the electron and ion spectrums in the $-100\leq x^{*}-\beta_{\rm c0}t^{*}\leq0$ region. The initial drifting population is visible around $p^{*}=10$. The cloud and ambient electrons are entirely mixed by $t^{*}=150$ and form a single electron population (Figure \ref{phase_space2} at $t^{*}=150$). By end of the simulation, however, the cloud and ambient ions are still distinguishable (bottom right panel in Figure \ref{phase_space2}). The energy exchange between the ion and electron is obvious in the over-plotted lines in Figure \ref{phase_space2} where ion spectrum gradually becomes wider, accompanied by decreasing number of ions at $p^{*}=10$. As one can see, the extension is asymmetric where the left side of the spectrum is wider than the right side because the kinetic energy of ions is consumed for acceleration of the electrons.  There is no sign of non-thermal electron acceleration because its spectrum is Maxwellian. In fact, the non-thermal power-law tail in the electron spectrum would appear beyond the energy of the order of the kinetic energy of cloud ions. Therefore, It occurs at much larger scale and longer timescale for this kind of simulations.

\begin{figure}
\includegraphics[scale=0.35]{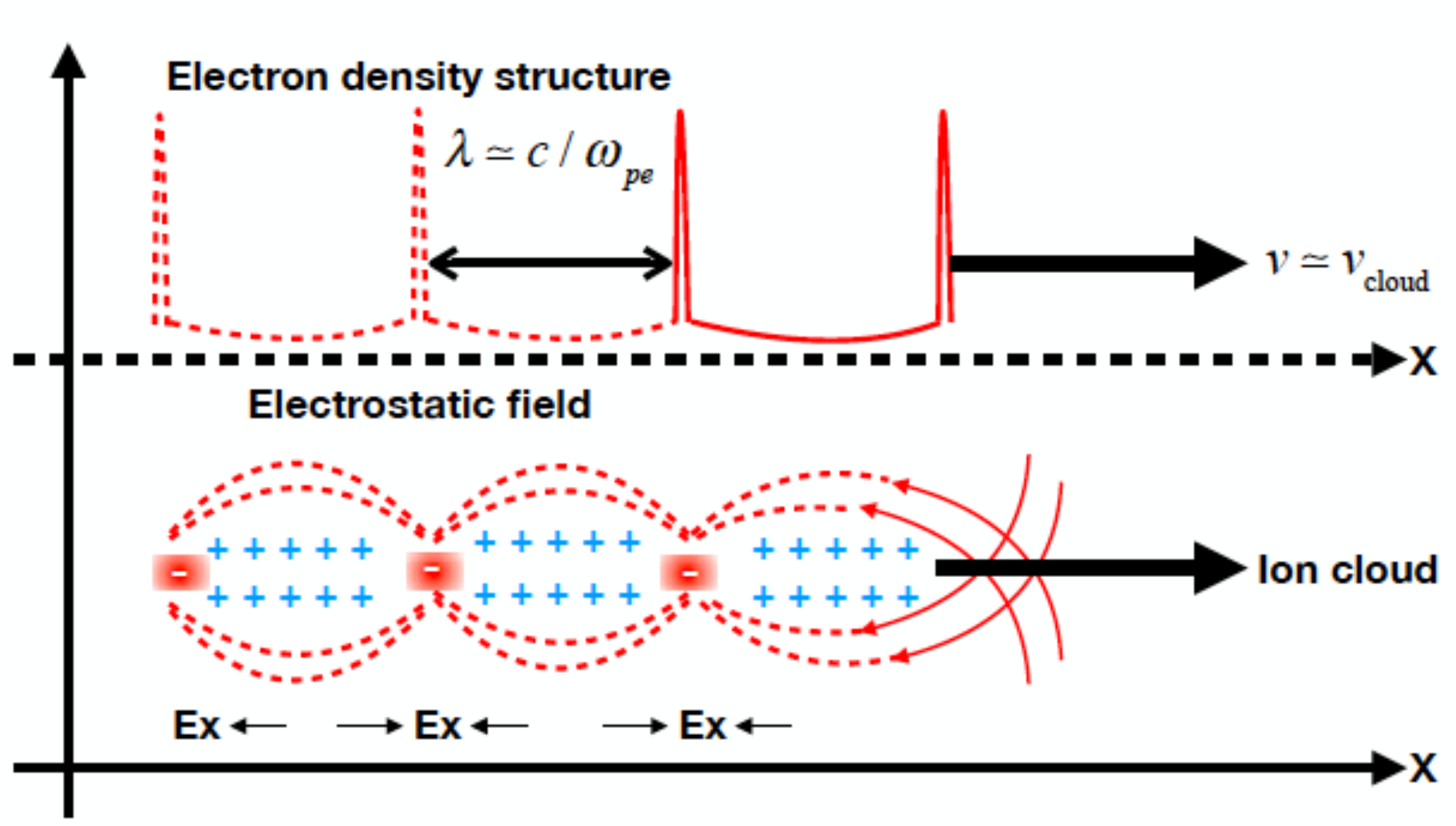}
\caption{The schematic diagram of the electron acceleration process within the $-100\leq x^{*}-\beta_{\rm c0}t^{*}\leq0$ region.}
\label{PWFA}
\end{figure}

\begin{figure}
\includegraphics[scale=0.4]{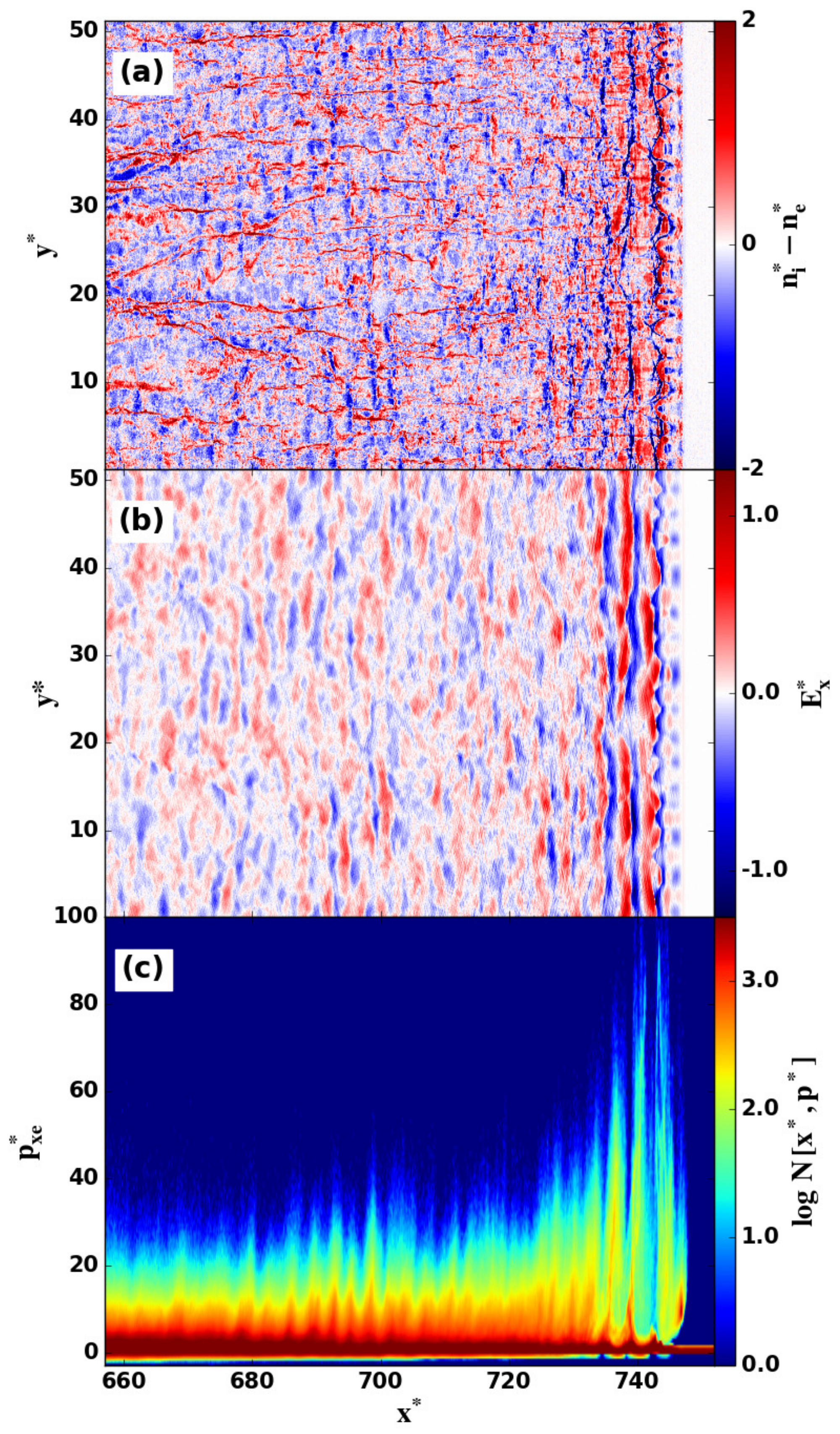}
\caption{The electron acceleration within the $-100\leq x^{*}-\beta_{\rm c0}t^{*}\leq0$ region. Shown are (a) particle density, red color for ion and blue color for electron, (b) ES filed, and (c) phase-space distribution of electrons.  }
\label{PWFA2}
\end{figure}

The acceleration of the electron within the $-100\leq x^{*}-\beta_{\rm c0}t^{*}\leq0$ region can be explained based in Figure \ref{PWFA}. Propagation of the ion cloud into the ambient plasma generates an accelerating structure which co-moves with the ion cloud. Following the ion cloud are alternating zones of high and low electron density that create a longitudinal accelerating electric field.  The ion cloud penetrates the ambient plasma and attracts the plasma electrons towards the axis of cloud propagation. The electrons attain kinetic energy as they accelerate towards the ion cloud and overshoot the axis, leaving a positively charged region of ions behind the cloud. They pinch together behind the propagating ion cloud as they rush back in to fill the positively charged zone. This creates a strong accelerating electric field in the direction of the ion cloud. Figure \ref{PWFA2} is representative of the characteristics discussed above. The figure shows the modulation of the ambient plasma by cloud ion in panel (a), the generated ES fields in panel (b), and the electron resonant acceleration in panel (c), respectively.

The amplitude of the ES wave can be approximated by balancing the wave energy density to the kinetic energy density of the ion cloud multiplied by a conversion factor.

\begin{equation} \label {dispersion}
\frac{E^{*2}}{2}=\zeta(\Gamma_{\rm {c0}}-1) R
\end{equation}

where $\zeta$ is the conversion factor of ion kinetic energy to the ES wave energy. The conversion factor can be calculated from the growth rate of the ES filed as discussed in Section \ref{Linear analysis}, $\zeta\simeq\frac{\sqrt{3}}{\gamma_0}(\frac{n^{*}_{\rm c0}}{2R})^{1/3}$ which reduce to $\zeta\sim (\frac{1}{R})^{1/3}$ in the no-relativistic regime. In the current work, we have used $\Gamma_{\rm {c0}}=10$ and $R=25$. Therefore, $\zeta\sim0.01$, and $\vert E^{*}\vert\sim1.5$ which is compatible with ES field amplitude shown in Figure \ref{PWFA2}b.

Other types of acceleration are also present in this kind of simulations. Behind the shocked region, as can be seen in Figure \ref{phase_space}c, there is a shock reflected population of the ambient ions with non-relativistic speed. The interaction of this population with the incoming cloud electrons forms a double layer plasma which is an efficient electron accelerator (Figure \ref{phase_space}b around $x^{*}\sim300$). The double layer plasma decelerates the ions and at the same time accelerate the electrons as shown in \citet{ard16}.

\section{Summary and conclusions}\label{Summary and conclusions}
The aim of this paper is an investigation of the possible processes that pre-accelerate electrons in the relativistic unmagnetized cloud-ambient interaction. The pre-acceleration processes increase the kinetic energy of electrons up to energy equipartition between electrons and ions, so that capable them for participating in the DSA at later stage. 

Although some mechanisms, e.g., SDA and ESA, are already proposed for pre-acceleration of the electrons in magnetized clouds, it is not fully understood for unmagnetized cases. We have used PIC simulation to shed light on the electron pre-acceleration for such kind of systems. In our simulation, an unmagnetized electron-ion cloud propagates into an ambient plasma in which a shock structure forms at later times.

The plasma instabilities are EM in the relativistic case, and electrons can be effectively heated in interaction with EM waves. 
We have shown that an oblique instability including both longitudinal ES, and transversal EM wave components dominates the unstable spectrum in the early stages. 
Because of heating, a thermal anisotropy arises and the instability turns to a Weibel-like instability afterward. We have seen that the transverse EM fields heated the species as $p^{*}_{\rm yi} \simeq p^{*}_{\rm ye}/R$ is preserved.

Additionally, there is an oblique instability where the origin of this instability is the propagation of the cloud ions into the ambient plasma. The ES component of this oblige mode is more pronounced compered to the EM one. The electron acceleration by the ES field is accompanied by deceleration of the cloud ions. In fact, it pumps the kinetic energy of the cloud ions to electrons with an efficiency of $\simeq\frac{\sqrt{3}}{\gamma_0}(\frac{n^{*}_{\rm c0}}{2R})^{1/3}$. The wavelength of the electron oscillation is the same as the wavelength of excited ES field. Hence, we have observed resonant electron acceleration.

No power-law population is seen in the electron spectrum. This work focuses on the early stage of the shock formation where electron and ion are still far from equilibrium. We expect a power-law distribution for a larger simulation box and on a longer timescale. 

We have used an ion-electron mass ratio of $m_{\rm i}/m_{\rm e}=25$ in the present work.  This low mass ratio is required to maintain the computational expenses of simulations feasible. However, it adjusts the growth rate of the instabilities as well. In the first growth stage $0\lesssim t^{*}\lesssim40$ , when the ions still are not involved in the instabilities, the energy of the magnetic fields grows exponentially, independent of the mass ratio. The mass ratio effect becomes important afterward. When it is small compared to the realistic one (1836), the saturation levels of the EM and ES fields become higher. Increasing the mass ratio will reduce the ion isotropization rate and the rate of kinetic energy exchange with electrons via the Weibel-like instabilities. The efficiency of energy exchange between the cloud ions and electrons is $\simeq\frac{\sqrt{3}}{\gamma_0}(\frac{n^{*}_{\rm c0}}{2R})^{1/3}$. We might reach an energy equipartition between electrons and ions for the mass ratio of $m_{\rm i}/m_{\rm e}=25$,  but it would not be achievable for the realistic mass ratio. Moreover, it is found that Weibel-like modes govern the high beam density regimes in the beam-plasma interactions \citep{brd10}. The domain of these modes expands as the mass ratio decreases. Consequently, the domains governed by the oblique modes shrink with decreasing the mass ratio. Therefore, our low mass ratio gives a higher importance to the Weibel-like instabilities than what they normally have.

\begin{acknowledgments}
The simulations presented here were performed on the KDK computer system at Research Institute for Sustainable Humanosphere, Kyoto University.
\end{acknowledgments}

\appendix
\section{Tensor elements for water-bag distribution function}\label{Tensor elements for water-bag distribution functions}
The $D_{\rm mn}$ elements for water-bag distribution function given at Eq. \ref{PDF} are presented in this section. Some parameters are used as follows: 

\begin{subequations} \label {parameters}
\begin{align}
p^{*}_{\pm}=p^{*}_0\pm p^{*}_{\parallel}\\
(\gamma_{\perp}, \beta_{\perp})=(\sqrt{1+p^{*2}_{\perp}}, p^{*}_{\perp}/\gamma_{\perp})\\
(\gamma_{\pm}, \beta_{\pm})=(\sqrt{1+p^{*2}_{\pm}}, p^{*}_{\pm}/\gamma_{\pm})\\
(\phi_{\pm}, \alpha_{\pm})=(\pm\beta_{\perp}\gamma_{\perp}/\gamma_{-},\pm\beta_{\perp}\gamma_{\perp}/\gamma_{+})\\
\theta_{\pm}=\beta_{\pm}\gamma_{\pm}/\gamma_{\perp}
\end{align}
\end{subequations}

The resulting $D_{\rm mn}$ reads

\begin{equation} \label {elements}
D_{\rm mn}=\delta_{\rm mn}+{\sum_{\rm{s}}}\frac{1}{R\omega^{*2}}(Z_{\rm mn}+Y_{\rm mn}k^{*}_{\rm y}+X_{\rm mn}k^{*}_{\rm x})
\end{equation}

where

\begin{subequations} \label {parameters}
\begin{eqnarray}
Z_{\rm yy}=-\frac{1}{2p^{*}_{\parallel}}[\ln{\frac{ p^{*}_{+} + \sqrt{1+p^{*2}_{+}+ p^{*2}_{\perp}} }{p^{*}_{-}  + \sqrt{1+p^{*2}_{-}+ p^{*2}_{\perp}}}}]
\end{eqnarray}

\begin{eqnarray}
Z_{\rm xy}=Z_{\rm yx}=0
\end{eqnarray}

\begin{eqnarray}
Z_{\rm xx}=\frac{1}{4p^{*}_{\parallel}p^{*}_{\perp}}[p^{*}_{-}\ln{\frac{ p^{*}_{\perp}  + \sqrt{1+p^{*2}_{-}+ p^{*2}_{\perp}} }{-p^{*}_{\perp}  + \sqrt{1+p^{*2}_{-}+ p^{*2}_{\perp}}}}\nonumber\\
 -p^{*}_{+}\ln{\frac{ p^{*}_{\perp}  + \sqrt{1+p^{*2}_{+}+ p^{*2}_{\perp}} }{-p^{*}_{\perp}  + \sqrt{1+p^{*2}_{+}+ p^{*2}_{\perp}}}}]
\end{eqnarray}

\begin{eqnarray}
Y_{\rm yy}=\frac{\beta_{\perp}}{4p^{*}_{\parallel}}[ I_0(\omega^{*},k^{*}_{\rm y}\beta_{\perp},-k^{*}_{\rm x}) \nonumber\\ -I_0(\omega^{*},-k^{*}_{\rm y}\beta_{\perp},-k^{*}_{\rm x})]^{\theta_{+}}_{\theta_{-}}
\end{eqnarray}

\begin{eqnarray}
Y_{\rm xy}=Y_{\rm yx}=-\frac{1}{4p^{*}_{\parallel}}[ I_1(\omega^{*},k^{*}_{\rm y}\beta_{\perp},-k^{*}_{\rm x}) \nonumber\\
+I_1(\omega^{*},-k^{*}_{\rm y}\beta_{\perp},-k^{*}_{\rm x})]^{\theta_{+}}_{\theta_{-}}
\end{eqnarray}

\begin{eqnarray}
Y_{\rm xx}=\frac{1}{4p^{*}_{\parallel}\beta_{\perp}}[ I_2(\omega^{*},k^{*}_{\rm y}\beta_{\perp},-k^{*}_{\rm x}) \nonumber\\
-I_2(\omega^{*},-k^{*}_{\rm y}\beta_{\perp},-k^{*}_{\rm x})]^{\theta_{+}}_{\theta_{-}}
\end{eqnarray}

\begin{widetext}

\begin{equation}
X_{\rm yy}=\frac{1}{4p^{*}_{\parallel}p^{*}_{\perp}}\{[\gamma_{-}I_2(\omega^{*},-k^{*}_{\rm x}\beta_{-},-k^{*}_{\rm y})] ^{\phi_{+}}_{\phi_{-}}
-[\gamma_{+}I_2(\omega^{*},-k^{*}_{\rm x}\beta_{+},-k^{*}_{\rm y})]^{\alpha_{+}}_{\alpha_{-}}\}
\end{equation}

\begin{equation}
X_{\rm xy}=X_{\rm yx}=\frac{1}{4p^{*}_{\parallel}p^{*}_{\perp}}\{[ p^{*}_{-} I_1(\omega^{*},-k^{*}_{\rm x}\beta_{-},-k^{*}_{\rm y})]^{\phi_{+}}_{\phi_{-}}
- [p^{*}_{+}  I_1(\omega^{*},-k^{*}_{\rm x}\beta_{+},-k^{*}_{\rm y})]^{\alpha_{+}}_{\alpha_{-}}\}
\end{equation}

\begin{equation}
X_{\rm xx}=\frac{1}{4p^{*}_{\parallel}p^{*}_{\perp}}\{[ p^{*}_{-}\gamma_{-}I_0(\omega^{*},-k^{*}_{\rm x}\beta_{-},-k^{*}_{\rm y})] ^{\phi_{+}}_{\phi_{-}}
- [p^{*}_{+}\gamma_{+}I_0(\omega^{*},-k^{*}_{\rm x}\beta_{+},-k^{*}_{\rm y})]^{\alpha_{+}}_{\alpha_{-}}\}
\end{equation}

\end{widetext}
\end{subequations}
The $I_{\rm n}(\phi, c_1, c_2, c_3)$  integrals which have analytical solutions are given by:
\begin{equation}
I_{\rm n}(\phi, c_1, c_2, c_3)=\int {d\phi \frac{\tan^{\rm n}\phi}{c_1+c_2\cos\phi + c_3\sin\phi} }
\end{equation}

\section{Numerical Cherenkov radiation}\label{numerical Cherenkov radiation}

The numerical dispersion relation of the electromagnetic waves using the implicit field solver reads:

\begin{equation} \label {FDTDdispersion}
[\frac{1}{c\Delta_{\rm t}}\tan{\frac{\omega \Delta_{\rm t}}{2}}]^2 =[\frac{1}{\Delta}\sin{\frac{k_{x} \Delta}{2}}]^2 + [\frac{1}{\Delta}\sin{\frac{k_{y} \Delta}{2}}]^2
\end{equation}

where $\Delta_{\rm x}=\Delta_{\rm y}=\Delta$ and the implicit parameter $\theta=0.5$. In our simulation, the cloud plasma travels in the $x$-direction with bulk velocity $\beta_{\rm c0}=0.995$. The dispersion relation related with movement of the plasma is therefore $\omega = \beta_{\rm c0} k_{\rm x}$. Numerical Cherenkov radiation happens if plasma flow passes the electromagnetic waves. This appears within wavenumber regions where the plasma cloud mode intersects the electromagnetic wave. The intersections might be obtained by substituting $\omega = \beta_{\rm c0} k_{\rm x}$ into Eq. \ref{FDTDdispersion}. Hence, the numerical Cherenkov radiation is supposed to be dominant at the wavenumbers given by: 

\begin{equation} \label {FDTDdispersion2}
 k_{y}=\frac{2}{ \Delta}\sin^{-1}\sqrt{{\rm {CFL}^{-2}}\tan^2{\frac{\beta_{\rm c0} k_{x} \Delta_{\rm t}}{2}} -\sin^2{\frac{k_{x} \Delta}{2}} }
\end{equation}

The performed PIC simulations using pCANS code are tested against Eq. \ref{FDTDdispersion2}.

\nocite{*}
\bibliography{manuscript}

\end{document}